\documentclass[12pt]{article}

\usepackage{amsmath,amssymb,theorem,amsmath,amsfonts,array,color}
\usepackage{psfrag}

\def\V0{{\bf V^0}}
\def\VFAC{{\bf V^0(FAC^0)}} 
\def\FAC{{\bf FAC^0}}
\def\SB{{\Sigma_0^B}}
\def\AC0{{\bf AC^0}}
\def\NN{{\mathbb{N}}}

\def\Yvec{\vec{Y}}
\def\Zvec{\vec{Z}}
\def\xvec{\vec{x}}

\def\lra{\leftrightarrow}

\def\V0{V^0}
\def\FAC{{\mathrm{FAC}^0}}
\def\AC0{{\mathrm{AC}^0}}
\def\VFAC{\V0 (\FAC )}
\def\TC0{\mathrm{TC}^0}

\def\Yvec{\bar{Y}}
\def\Zvec{\bar{Z}}
\def\xvec{\bar{x}}

\def\N{\mathbb N}

\let\implies=\rightarrow

\let\proves=\vdash
\let\epsilon=\varepsilon

\def\Proof{\noindent{\bf Proof}\quad}

\newcommand\E[1]{\exists #1 \,}
\newcommand\A[1]{\forall #1 \,}

\def\ss{\subseteq}

\def\P{\mathrm{P}}

\def\PHP{\mathrm{PHP}}
\def\WPHP{\mathrm{WPHP}}

\def\BB{\mathrm{BB}}
\def\PV{\mathrm{PV}}
\def\QPV{\mathrm{QPV}}
\def\ID0{\mathrm{I \Delta_0}}

\def\BASIC{\mathrm{BASIC}}
\def\LIND{\mathrm{LIND}}
\def\IND{\mathrm{IND}}

\def\S{\Sigma}

\def\bx{\bar{x}}

\def\bb{\bar{b}}
\def\bc{\bar{c}}

\def\bz{\bar{z}}

\def\DCR{\Delta^b_1-\mathrm{CR}}
\def\VTC{\mathrm{VTC}^0}
\def\Sb{\Sigma^b}  
\def\Pb{\Pi^b}
\def\Db{\Delta^b}

\def\EPV{\exists^b \PV}

\newcommand{\ang}[1] { \langle #1 \rangle }

\def\bi{ \{ 0,1 \} }
\def\N{\mathbb{N}}
\def\<={\leqslant}
\def\>={\geqslant}
\def\conj{\wedge}
\def\disj{\vee}
\def\implies{\rightarrow}

\def\QED{\nolinebreak \hspace*{\fill} \ensuremath{\square}}
\def\line{\vspace{12pt}}

\newcommand{\es}[2]{\exists #1 \! < \! #2 \,}
\newcommand{\as}[2]{\forall #1 \! < \! #2 \,}

\title{The strength of replacement in weak arithmetic}
\author{Stephen Cook and Neil Thapen}

\newtheorem{theo}{Theorem} 

\newtheorem{lemma}[theo]{Lemma}
\newtheorem{cor}[theo]{Corollary}

\begin{document}
\title{The strength of replacement in weak arithmetic}

\author{Stephen Cook and Neil Thapen\\
Department of Computer Science, University of Toronto \\
\{sacook, thapen\}@cs.toronto.edu\\
}

\maketitle
\thispagestyle{empty}

\begin{abstract}

The {\em replacement} (or {\em collection} or {\em choice})
axiom scheme $\BB(\Gamma)$ asserts bounded quantifier exchange
as follows:
\[
\as{i}{|a|}\es{x}{a}\phi(i,x)
 \implies \E w \as{i}{|a|} \phi(i,[w]_i)
\]
where $\phi$ is in the class $\Gamma$ of formulas.  The theory $S^1_2$
proves the scheme $\BB(\Sb_1)$, and thus in $S^1_2$ every $\Sb_1$
formula is equivalent to a strict $\Sb_1$ formula (in which all
non-sharply-bounded quantifiers are in front).  Here we prove
(sometimes subject to an assumption) that
certain theories weaker than $S_2^1$ do not prove either $\BB(\Sb_1)$
or $\BB(\Sb_0)$.
We show (unconditionally) that $\V0$ does not prove $\BB(\SB)$, where
$\V0$ (essentially I$\S^{1,b}_0$) is the two-sorted theory associated
with the complexity class AC$^0$.  We show that PV does not prove
$\BB(\Sb_0)$, assuming that integer factoring is not possible in
probabilistic polynomial time. 
Johannsen and Pollet introduced the theory $C^0_2$ associated with
the complexity class TC$^0$, and later introduced an apparently
weaker theory $\DCR$ for the same class. 
We use our methods to show that $\DCR$ is indeed weaker
than $C^0_2$, assuming
that RSA is secure against probabilistic
polynomial time attack.  

Our main tool is the KPT witnessing theorem.
\end{abstract}

\section{Introduction}

We are concerned with the strength of
various theories of bounded arithmetic
associated with the complexity classes P, TC$^0$, and AC$^0$.
Our goal is to show that some of these theories cannot prove
replacement, which is the axiom scheme 
\begin{equation}\label{repl}
\as{i}{|a|}\es{x}{a}\phi(i,x) 
 \implies \E w \as{i}{|a|} \phi(i,[w]_i).
\end{equation}
(where $\phi(i,x)$ can have other free variables).
We use $\BB(\Gamma)$ to denote replacement
for all formulas $\phi$ in a class $\Gamma$ (usually $\Sb_0$ or
$\Sb_1$).
Replacement is also sometimes known as 
``collection'' (eg. \cite{krajicek})
 or ``choice'' (eg. \cite{zambella}).
We begin by briefly describing the main theories of interest.

The language of first order arithmetic that we use is
\[
\{ 0,1,+,\cdot, <, |x|, (x)_i, [x]_i, x \# y \}.
\]
Here 
$|x|$ is the length of $x$ in binary notation, $(x)_i$ is the
$i$th bit of $x$, $[x]_i$ is the $i$th element of the sequence
coded by $x$, and $x \# y$ is $2^{|x| \cdot |y|}$.
All our theories in this language are assumed to include 
a set of axioms $\BASIC$ fixing the algebraic properties of these
symbols; see \cite{buss,krajicek} for more detail.

In the first order setting we will look at $\BB(\S^b_0)$, 
or ``sharply bounded replacement''.
A sharply bounded or $\S^b_0$ formula is one in which every
quantifier is bounded by a term of the form $|t|$. A $\Sb_1$ 
formula is a sharply bounded formula preceded by a mixture
of bounded existential and sharply bounded universal quantifiers. 
A strict $\Sb_1$ formula is a sharply bounded formula preceded
by a block of bounded existential quantifiers.

The strongest theory we look at is $S^1_2$ \cite{buss}, defined as
 $\BASIC$ together with ``length induction'', that is the $\LIND$ axiom
\begin{equation}\label{LIND}
\phi(0) \conj \as{x}{|a|}(\phi(x) \implies \phi(x+1))
\implies \phi(|a|)
\end{equation}
for all $\Sb_1$ formulas $\phi$. 

$S^1_2$ proves $\BB(\Sb_1)$, and hence for every $S^1_2$-formula
$\phi$ there is a strict-$\Sb_1$ formula $\phi'$ such that
$S^1_2$ proves $(\phi \lra \phi')$.  This fact may have influenced
Buss's \cite{buss} original decision not to choose strict $\Sb_i$
as the standard definition of $\Sb_i$.  The general definition
allows Buss to prove \cite{buss} Thm 2.2 showing that if a theory $T^+$
extends $T$ by adding $\Sb_1$-defined function symbols then
$\Sb_1$ formulas in the extended language are provably equivalent
to $\Sb_1$formulas in the original language.   This result may not
hold if $\Sb_1$ is taken to be strict $\Sb_1$ and $T$ does not
prove replacement.  We show here that certain weaker theories (likely) do
not prove replacement.  For these theories, strict $\Sb_1$ is
a more appropriate definition, and extensions by $\Sb_1$-defined
functions must be handled with care.

The first order theory we will use most often is
$\PV$ \cite{cookPV} (called $\PV_1$ in \cite{krajicek} and $\QPV$
in \cite{cook98}). This is defined
by expanding our language to include a function symbol for every 
polynomial time algorithm, introduced inductively by Cobham's limited
recursion on notation. These are called $\PV$ functions, and 
quantifier free formulas in this language are $\PV$ formulas.
One way to axiomatize $\PV$ is $\BASIC$ plus
universal axioms defining the new function symbols plus the
induction scheme $\IND$
\[
\phi(0) \conj \as{x}{a}(\phi(x) \implies \phi(x+1))
\implies \phi(a)
\]
for open formulas $\phi(x)$.  However it is an important fact that $\PV$
is a universal theory, and can be axiomatized by its universal
consequences \cite{buss,cook98}.

$\PV$ and $S^1_2$ are closely linked to the complexity class $\P$.
The provably total $\Sb_1$ (or even strict $\Sb_1$) functions in these
theories are precisely the polynomial time functions.
$S^1_2$ is $\Sb_1$-conservative over $\PV$ \cite{buss},
but $\PV$ cannot prove the $\Sb_1$-$\LIND$ axiom scheme (\ref{LIND})
for $S^1_2$ unless the polynomial hierarchy (provably) collapses
\cite{KPT,busshierarchy,zambella}.

First order theories are unsuitable for dealing with very weak 
complexity classes such as $\AC0$, in which we cannot even
define multiplication of strings. 
In this setting it is more natural to work with a two-sorted
or ``second order'' theory.
$\V0$ is the 
theory described in the Notes \cite{cook02},
page 56.  It is based on $\Sigma^p_0$-comp \cite{zambella}
and is essentially the same as ${\mathrm I}\Sigma^{1,b}_0$.
The two sorts are numbers and strings (finite sets of numbers).
The axioms consist of number axioms giving the basic properties
of $0,1,+,\cdot,\leq$, two axioms defining the ``length'' $|X|$
of a finite set $X$ to be 1 plus the largest element in $X$,
or 0 if $X$ is empty, and the comprehension scheme for $\SB$
formulas.  The $\SB$ formulas allow bounded number quantifiers, but
no string quantifiers, and represent precisely the uniform $\AC0$ relations
on their free string variables.

If we add to $V^0$ a function $X \cdot Y$ for string multiplication,
we get a theory equivalent to the first order theory $\Sb_0-\LIND$.
The number sort would correspond to sharply bounded numbers and the
string sort to ``large'' numbers; the $\SB$ induction available in
$V^0$ would correspond to $\Sb_0-\LIND$.

With this correspondence (known as RSUV isomorphism \cite{takeuti93,raz93})
in mind, we consider $V^0$ 
and the first order fragments of $S^1_2$
as fitting naturally
into one hierarchy of theories of bounded
arithmetic. 
The only differences between the two approaches will be in
the notation for strings and sequences. $(z)_i=1$ in the first
order setting corresponds to $Z(i)$ or $i \in Z$ in the second
order setting; $[z]_i$ corresponds to $Z^{[i]}$ (see next paragraph).

In second order bounded arithmetic the replacement scheme (\ref{repl})
becomes
\[
\as{i}{n}\es{X}{n} \phi(i,X)
 \implies \E W \as{i}{n} \phi(i,W^{[i]}).
\]
Here $\es{X}{n} \phi$ stands for $\exists X (|X|<n\wedge \phi)$ and
$W^{[i]}(u)$ is formally $W(\langle i,u\rangle)$
where $\langle i,u\rangle$ is a standard pairing function
(so $W^{[i]}$ is row $i$ in the two-dimensional bit array $W$).

\begin{figure}
\input{diagram.pstex_t}
\end{figure}

Our main results are that $\V0$ does not prove $\Sigma_0^B$ replacement
(unconditionally) and that, unless integer
factoring is possible in probabilistic
polynomial time, $\PV$ does not prove $\Sb_0$ replacement. 
(As mentioned above, $S^1_2$ does prove $\Sb_0$ replacement.)

We summarize our results with a picture of the structure of theories
between $S^1_2$ and $V_0$. An arrow on the diagram represents inclusion.
To the right of an arrow we give a sufficient condition for the two
theories to be distinct. 
A bold arrow indicates that this condition is true, and that the 
theories in fact are distinct.
To the left of an arrow we show the 
conservativity between the two theories.

We will begin with the bottom of the diagram.
We have already talked about $\V0$ and $\PV$. 
$\DCR$ was introduced by Johannsen and Pollett in 
\cite{Db1-CR} to correspond to the complexity class 
$\TC0$ of constant-depth circuits with threshold gates. The
$\Sb_1$ functions provably total in $\DCR$ are precisely
the uniform $\TC0$ functions. The theory is defined as the closure
of the $\BASIC$ axioms and the $\LIND$ axioms for open formulas
 under the normal
rules of logical deduction together with the $\Db_1$-comprehension rule:
if we can {\em prove} that a $\Sb_1$ formula $\phi(x)$ is equivalent
to a $\Pb_1$ formula $\psi(x)$, then are allowed to introduce comprehension 
for $\phi$,
\[
\E w \as{i}{|a|}, (w)_i = 1 \leftrightarrow \phi(i).
\]

$\DCR$ proves induction for sharply bounded formulas, so 
we can think of $V^0$ as a subtheory of it.  In fact \cite{nguyen}
defines an extension $\VTC$ of $V^0$ by adding an axiom for the
function NUMONES$(X)$ (which counts the number of 1's in the string $X$)
and proves $\VTC$ is RSUV isomorphic to $\DCR$.  But $\VTC$ proves
the pigeonhole principle, as represented by a $\SB$ formula $\PHP(X,n)$
\cite{nguyen}, and $V^0$ does not \cite{cook02}. 
Hence $\DCR$ is strictly stronger than $\V0$.

The $\Db_1$-comprehension rule is a derived rule of $\PV$.
This is because by results in \cite{buss} if a formula $\phi$
is provably $\Db_1$ in $\PV$, then $\PV$ proves that the
characteristic function of $\phi$ is computable in polynomial time,
and hence that comprehension holds for $\phi$.  Thus $\PV$
is an extension of $\DCR$.

$\PV$ is separated from $\DCR$ by the circuit value principle,
which says that ``for all circuits $C$ and all inputs $\bx$, there
exists a computation of $C$ on $\bx$''. This is provable in $\PV$, 
but under the assumption that $\P$ does not equal uniform $\TC0$
it is not provable in $\DCR$.

Turning now to the top of the diagram, \cite{buss} proves
the $\A \S^b_1$-conservativity of $S^1_2$ over $\PV$. 
If $\PV + \BB(\Sb_0)$ proves
$S^1_2$, then $\PV \proves S^1_2$ \cite{zambella}
and hence the bounded arithmetic hierarchy collapses to $\PV$ 
and the polynomial hierarchy PH collapses to $\S^p_2\cap\Pi^p_2$
\cite{zambella,busshierarchy}.

The $\forall \exists \S^B_0$-conservativity of 
$V^0+\BB(\S^B_0)$ over $V^0$ 
is from Zambella \cite{zambella}.
$\S^b_0-\LIND + \BB(\S^b_0)$ was introduced in \cite{C02}
by Johannsen and Pollett (where they call it $C^0_2$), 
and proved to be
$\forall \Sb_1$ conservative over $\DCR$
in \cite{Db1-CR}.
From these conservativity results
it follows that $V^0+\BB(\S^B_0)$ does not prove the 
pigeonhole principle and $\DCR+\BB(\S^b_0)$ does 
not prove the circuit value principle (unless $\P$ equals uniform $\TC0$), 
which gives us the separations between the three theories with replacement.

In the body of the paper we show the separations between the 
theories with and without various kinds of replacement, using a similar
argument in all cases. 

In section 2 we describe how our general argument goes.
In section 3 we use it together with the fact that parity is not
computable in nonuniform $\AC0$ to separate $\V0$ from $\V0+\BB(\Sb_0)$.

In section 4 we
show that if $\PV$ proves $\S^b_0$-replacement, then factoring
is possible in probabilistic polynomial time.  (This strengthens
a result in \cite{thapen} where the weaker conclusion ``RSA is
insecure'' was proved.)
We observe that this
is true even if we look at weak versions of $\S^b_0$-replacement,
where we code very short sequences of witnesses; for example
$\BB(\S^b_0,||x||)$ in the diagram is the scheme of
replacement for sequences of double-log length:
\[
\as{i}{||a||}\es{y}{a}\phi(i,y) 
 \implies \E w \as{i}{||a||} \phi(i,[w]_i).
\]
The dotted line in the diagram represents the fact that if factoring
is hard, then all the theories 
$\BB(\S^b_0,|x|)$,
$\BB(\S^b_0,||x||)$,
$\BB(\S^b_0,|||x|||)$, $\ldots$ are distinct (in fact we show
something slightly stronger than this).
By a similar argument, all these theories are distinct over
$\V0$ (in place of $\PV$), without any assumptions, but for the
sake of tidiness we have not put this on the diagram.

The theory of  
strong $\Db_1$ comprehension
is like $\DCR$, except that rather than having a rule
that if a formula is provably $\Db_1$ then comprehension
holds for it, we have the ``$\Db_1$ comprehension axiom scheme''
\begin{equation}\label{compre}
\A x (\phi(x) \leftrightarrow \neg \psi(x))
 \implies \E w \as{i}{|a|} (\phi(i) \leftrightarrow (w)_i=1)
\end{equation}  
where $\phi, \psi \in \Sb_1$ (and may contain other parameters); 
so comprehension holds for 
$\phi$ in a structure, if $\phi$ is $\Db_1$ in that structure.
The question is raised in \cite{Db1-CR}, whether
this theory is strictly stronger than $\DCR$.
We show that it is, under a cryptographic assumption.
We consider a principle not shown on the diagram, 
which we call ``unique replacement''. We show that if RSA is secure
against probabilistic polynomial time attack then $\PV$ does not
prove unique replacement, and that it follows that $\PV$, and hence
$\DCR$, does not prove the $\Db_1$ comprehension axiom scheme.

We have not looked for a separation between this last theory
and $\S^b_0-\LIND + \BB(\S^b_0)$.

A preliminary version of this paper appears in \cite{cook_thapen}.

\section{Witnessing with an interactive computation}

First we recall a standard lemma. 

\begin{lemma}
Over $\BASIC$, $\Sb_0$-replacement is equivalent to 
strict $\Sb_1$-replacement. Hence over $\PV$, $\Sb_0$-replacement
is equivalent to replacement for $\PV$ formulas, since $\PV$
proves that every $\PV$ formula is equivalent to a strict
$\Sb_1$ formula.

Similarly over $\V0$, $\SB$-replacement is
equivalent to $\S^B_1$-replacement, where a $\S^B_1$ formula
is a $\SB$ formula preceded by a block of bounded existential
string quantifiers.
\QED
\end{lemma}

Our main tool in this paper is the KPT witnessing theorem. 
We state it here for $\PV$, although it holds in a much
more general form.

\begin{theo} \cite{KPT}
Let $\phi$ be a $\PV$ formula and suppose $\PV \proves
\A x \E y \A z \phi(x,y,z)$.
Then there exists a finite sequence $f_1, \ldots, f_k$ of $\PV$ function
symbols such that 
\begin{align*}
\PV \proves \A x \A {\bz,}
& \phi(x,f_1(x),z_1) \disj \phi(x,f_2(x,z_1),z_2) \\
& \disj \ldots \disj \phi(x,f_k(x,z_1, \ldots, z_{k-1}),z_k).
\end{align*}
\end{theo}

\Proof
Let $b,c_1,c_2,...$ be a list of new constants, and let
$t_1,t_2,...$ be an enumeration of all terms built from symbols of
$\PV$ together with $b,c_1,c_2,...$, where the only new constants
in $t_k$ are among $\{b,c_1,...,c_{k-1}\}$.  It suffices to show that
\[
  \PV\cup \{\neg\phi(b,t_1,c_1),\neg\phi(b,t_2,c_2), \ldots ,
           \neg\phi(b,t_k,c_k)\}
\]
is unsatisfiable for some $k$.

Suppose otherwise.  Then by compactness
\begin{equation}\label{star}
  \PV\cup \{\neg\phi(b,t_1,c_1),\neg\phi(b,t_2,c_2),...\}
\end{equation}
has a model $M$.  Since $\PV$ is universal, the
substructure $M'$ consisting of the denotations of the
terms $t_1,t_2,...$ is also a model for (\ref{star}).  It is easy
to see that
\[
   M'\models \PV + \forall y\exists z\neg\phi(b,y,z)
\]
and hence $\PV\not\vdash \forall x\exists y\forall z\phi(x,y,z)$.
\QED
\line

Now choose a function $f$ which can be computed in 
polynomial time  
but which is hard to invert.
Suppose $\PV$ proves the following instance of replacement 
(which has $a$ and $y$ as parameters, and $m=|a|$):
\[
\as{i}{m} \es{u}{a} f(u)=[y]_i
\implies \E w \as{j}{m} f([w]_j)=[y]_j.
\]
We can rewrite this as
\[
\es{i}{m} \E w \as{u}{a}, \, f(u)=[y]_i 
\implies \as{j}{m} f([w]_j)=[y]_j. 
\]
Applying our witnessing theorem, we get $k \in N$ and functions
$g_1, \ldots, g_k$ and $h_1, \ldots h_k$ (which have $a$
as a suppressed argument), such that  
\begin{align*}
\PV &  \proves \as{\bz}{a},  \\
& (f(z_1)=[y]_{g_1(y)} \implies \as{j}{m} f([h_1(y)]_j)=[y]_j) \\
& \disj (f(z_2)=[y]_{g_2(y,z_1)} \implies 
 \as{j}{m} f([h_2(y,z_1)]_j)=[y]_j) \\
& \disj \ldots \\
& \disj (f(z_k)=[y]_{g_k(y,z_1, \ldots, z_{k-1})}  \implies \\
&  \qquad  \as{j}{m} f([h_k(y,z_1, \ldots, z_{k-1})]_j)=[y]_j)\\
\end{align*}
This allows us to write down an algorithm which given an input
$y$  (considered as a sequence $[y]_0, \ldots, [y]_{m-1}$) will
ask for a pre-image of $f$ on at most $k$ elements of $y$,
and with this information will output a number $w$
coding a sequence of pre-images of all $m$ elements of $y$.

The algorithm is as follows. 
Let $w=h_1(y)$. If  $\as{j}{m} f([w]_j)=[y]_j$ then
output $w$ and halt. Otherwise calculate $g_1(y)$ and ask
for a pre-image of $[y]_{g_1(y)}$; store the answer as $z_1$.
Then let $w=h_2(y,z_1)$. If  $\as{j}{m} f([w]_j)=[y]_j$ then
output $w$ and halt. Otherwise calculate $g_2(y,z_1)$ and ask
for a pre-image of $[y]_{g_2(y,z_1)}$; store the answer as $z_2$, and so on.
By our assumption the algorithm will run for at most $k$ steps
of this form before it outputs a suitable $w$.

Now fix $a$ such that $|a| = m > k$, and choose a
sequence $[x]_0, \ldots, [x]_{m-1}$ of numbers less than $a$. Let 
$y$ encode the pointwise image of $x$ under $f$. Run the algorithm
above, and reply to queries with elements of $x$. We will end up
with $w$ encoding a sequence of pre-images of $y$, which will clash
in some way with our assumption that $f$ is hard to invert.
 If $f$ is an injection,
$w$ will be the same as $x$; we use this in section 3. If $f$
is not an injection and $x$ was chosen at random, then $w$ is probably
different from $x$; we use this in sections 4 and 5.

The important properties of $\PV$ used in the argument above are
that it is universal and can define functions by cases (needed
for the KPT witnessing theorem) and that it can manipulate sequences. 
We show now how to make $V^0$ into a universal theory in which we
can carry out the same argument. 

We start by referring to
\cite{cook02}, pp 66--73.  A relation $R(\xvec,\Yvec)$ is in 
(uniform) $\AC0$
iff it is defined by some $\SB$ formula $A(\xvec,\Yvec)$.
A number function
 $f:\NN^k \times (\{0,1\}^\ast)^\ell \longrightarrow \NN  $
is an $\AC0$ function iff there is an $\AC0$ relation $R$ and a polynomial $p$
such that
\begin{equation}\label{numberF}
   f(\xvec,\Yvec) = \min z<p(\xvec,|\Yvec|)\ R(z,\xvec,\Yvec)
\end{equation}
A string function $F(\xvec,\Yvec)$ is an $\AC0$ function iff 
$|F(\xvec,\Yvec)|\leq p(\xvec,|\Yvec|)$ for some polynomial $p$, and
the bit graph
$$ B_F(i,\xvec,\Yvec)\equiv F(\xvec,\Yvec)(i) $$
is an $\AC0$ relation.

We denote by $\VFAC$ a conservative extension of $\V0$ obtained
by adding a set $\FAC$ of function symbols
with universal defining axioms for all
$\AC0$ functions, based on the above characterizations. 
$\FAC$ is essentially ${\cal R}-def$ in \cite{zambella}.)
This can be done
in such a way that $\VFAC$ is a universal theory.
In particular, the $\SB$ comprehension axioms follow since
for every $\SB$ formula $\phi$ there is a  $\FAC$ string function whose
range is the set of strings asserted to exist by the the comprehension
axiom for $\phi$.
Further, from (\ref{numberF}) it is clear that for every $\SB$ formula
$\phi$ there is a quantifier-free formula $\phi'$ in the language
of $\VFAC$ such that
$$\VFAC\vdash (\phi\lra\phi')  $$

From these remarks, it is clear that the usual proof of the KPT
witnessing theorem can be adapted to show the following:

\begin{theo}
Let $\phi(X,Y,Z)$ be a $\SB$ formula
such that 
$  \V0 \vdash \forall X\exists Y\forall Z\phi(X,Y,Z)  $.
Then there are $\FAC$ functions $F_1,...,F_k$ such that 
\begin{align*}\label{KPT}
\VFAC &  \proves  \forall X\forall \Zvec,  \\
& \phi(X,F_1(X),Z_1) \vee
   \phi(X,F_2(X,Z_1),Z_2) \\
& \vee \ldots \vee \phi(X,F_k(X,Z_1,...,Z_{k-1}),Z_k). 
\end{align*}
\end{theo}

Using this we can show that if $V^0$ proves $\SB$-replacement,
then for any $\AC0$ function $F$ there exists $k \in N$ and a uniform
$\AC0$ algorithm that will find a pre-image under $F$ of any sequence
$Y^{[0]}, \ldots, Y^{[m-1]}$ of strings by asking at most
$k$ queries of the form ``what is a pre-image of $Y^{[i]}$?''

\section{Replacement in $\mathbf{V^0}$ and parity}

Let $PARITY$ be the set
of all strings over $\{0,1\}$ with an odd number of 1s.
By a (nonuniform) $\AC0$ circuit family we mean a polynomial size bounded 
depth
family $\langle C_n:n\in\NN \rangle$
of Boolean circuits over $\wedge, \vee, \neg$
such that $C_n$ has $n$ inputs and one output.
Ajtai's theorem \cite{ajtai, FSS}
states that no such circuit family accepts $PARITY$.

We show that if $\V0$ proves the $\SB$ replacement scheme,
then (using KPT witnessing) there exists a (uniform) randomized $\AC0$
algorithm for $PARITY$.  This algorithm shows the existence of a (uniform)
$\AC0$ circuit family such that each circuit has a vector $\bar{r}$
of random input bits in addition to the standard input bits,
and with probability $p> 2/3$ the circuit correctly determines
whether the standard input is in $PARITY$ and with probability $1-p$
the circuit produces an output indicating failure.
From this a standard argument shows the existence of a nonuniform
$\AC0$ circuit family for parity, violating the above theorem.

Let $PAR$ be the function that maps a binary string of length $m$
to its parity vector. That is, $PAR(m,Y)=X$ if $|X|<m$ and,
for each $i<m$, $X(i)$ is the 
parity of the string $Y(0) \ldots Y(i)$.  In what follows
we take $m$ to be a parameter, assume $Y$ is an $m$-bit string,
and suppress the argument $m$ from $PAR(m,Y)$.

Plainly $PAR(Y)$ cannot be computed in $\AC0$. However its inverse,
which we will call $UNPAR$, is in uniform $\AC0$: 
the $i$th bit of $UNPAR(X)$
is given by the $\SB$ formula 
$(i=0 \conj X(i)) \disj (i > 0 \conj X(i-1) \oplus X(i) )$.
Here $UNPAR$ has an argument $m$, which we suppress.

Notice also that for all $m$-bit strings $A, B, C$, writing $\oplus$
for bitwise $XOR$, if
$A = B \oplus C$
then $PAR(A) = PAR(B) \oplus PAR(C)$.

\begin{theo}
$V^0$ does not prove $\BB(\SB)$.
\end{theo}

\Proof
Suppose $V^0 \proves \BB(\SB)$. Then
applying the argument of section 2 to the function $UNPAR$,
 for some fixed $k$ there is
a uniform $\AC0$ algorithm which, for any 
sequence $Y^{[0]}, \ldots, Y^{[m-1]}$ of binary strings of length $m$
makes $k$ queries of the form ``what is $PAR(Y^{[i]})$?'' and
outputs the sequence of parity vectors of $Y$.
 
We will show how to use this algorithm to compute the parity of a
single string in uniform randomized $\AC0$. Suppose $m \ge 3k$ and
let $I$ be the input string of length $m$ which we want to compute
the parity of.

Choose $m$ strings $U_0, \ldots, U_{m-1}$ in $\bi^m$ at random, and
for each $i$ compute $V_i = UNPAR(U_i)$. Choose a number $r$,
$0 \le r < m$, uniformly at random.
Define the string $Y$ (thought of as an $m \times m$ binary matrix)
by the condition
$$   Y^{[i]} = \left\{  \begin{array}{ll}
      V_i   & \mbox{if $i\not= r$}  \\
      I\oplus V_r  &  \mbox{if $i=r$.}
             \end{array}
   \right.   $$

Since for each $m$ the function $UNPAR$ defines a bijection from
the set $\{0,1\}^{m}$ to itself, and since for each $I$ with
$|I| < m$ the map $X \mapsto I\oplus X$ also defines a bijection
from  that set to itself, it follows that the string $Y$ defined
above, interpreted as an $m \times m$ bit matrix, is uniformly
distributed over all such matrices.

Now run our interactive $\AC0$ algorithm on $Y$. If the algorithm
queries ``what is $PAR(Y^{[i]})$?'' for $i \neq r$, reply with
$U_i$ (which is the correct answer).
If the algorithm queries ``what is $PAR(Y^{[r]})$?'', then abort
the computation.

Since at most $k$ different values of $i$ are compared to $r$ 
and since for each input $I$ each pair $(Y,r)$ is equally likely to
have been chosen, it follows that the computation will be aborted 
with probability at most $k/m\le 1/3$.

Hence  with probability at least 2/3 the algorithm is not
aborted, we are able to answer all the queries correctly, 
 and we obtain $W$ such
that $W^{[r]}=PAR(Y^{[r]})=PAR(I \oplus V_r)$.  But
$I = V_r\oplus (I\oplus V_r)$ and hence 
\begin{align*}
   PAR(I) &  =   PAR(V_r)\oplus PAR(I\oplus V_r) \\
            &  =   U_r  \oplus W^{[r]}
\end{align*}
We use this to compute $PAR(I)$ and use bit $m-1$
of $PAR(I)$
to determine whether $I\in PARITY$.  

For each input $I$
the algorithm succeeds
with probability at least 2/3, where the probability is taken over
its random input bits.

Since no such $\AC0$ algorithm exists, it follows that $\V0$
does not prove the $\SB$ replacement scheme. \QED

\section{Replacement in PV and factoring}

 We adapt the proof \cite{rabin}
that cracking Rabin's cryptosystem based on squaring
modulo $n$ is as hard as factoring.

Let $n$ be the product of distinct odd primes $p$ and $q$.
Suppose $0<x_1<n$ and $\gcd(x_1,n) = 1$. Let $c=x_1^2$. Then $c$ has 
precisely four square roots $x_1,x_2,x_3,x_4$
modulo $n$, as follows.
 
Let $x_p= (x_1 \bmod p)$ and $x_q = (x_1 \bmod q)$.
By the Chinese remainder theorem there are uniquely determined
numbers $x_1,x_2,x_3,x_4$ with $0<x_i<n$ such that
\[
\begin{array}{lll}
 x_1 \equiv x_p \pmod p & \qquad & x_1 \equiv x_q \pmod q   \\
 x_2 \equiv x_p \pmod p & \qquad & x_2 \equiv -x_q \pmod q   \\
 x_3 \equiv -x_p \pmod p & \qquad & x_3 \equiv x_q \pmod q   \\
 x_4 \equiv -x_p \pmod p & \qquad & x_4 \equiv -x_q \pmod q   
\end{array}
\]

Now $x_1-x_2 \equiv 0 \pmod p$ and $x_1-x_2 \equiv 2x_q \not \equiv 0 \pmod q$,
so $\gcd(x_1-x_2,n)=p$. So from $x_1$ and $x_2$ we can recover
$p$, and similarly from $x_1$ and $x_3$ we can recover $q$.

Hence if we have one square root of $c$, and are then given a square
root at random, we can factor $n$ with probability $1 \over 2$. 

\begin{theo} \label{the:roots}
If $\PV$ proves replacement for sharply bounded formulas,
then factoring (of products of two odd primes) is possible in 
probabilistic polynomial time.
\end{theo}

\Proof
We will use our standard argument, taking 
squaring modulo $n$ as our function $f$ (so 
$f$ has $n$ as a parameter).

If $\PV$ proves $\BB(\Sb_0)$
then there is polynomial time algorithm which, for
some fixed $k \in \N$, given any sequence
$y_0, \ldots, y_{m-1}$ of squares (modulo $n$),
makes at most $k$ queries of the form
``what is the square root of $y_i$?''
and, if these are answered correctly,
outputs square roots of all the $y_i$s.

Now suppose $n$ is large enough that $m = |n| > k$. 
Choose numbers $x_0, \ldots, x_{m-1}$ uniformly at random with
$0<x_i<n$. We may assume that $\gcd(x_i,n) = 1$ for all $i$,
since otherwise we can immediately find a factor of $n$.

For each $i$ let $y_i = (x_i^2 \bmod n)$. Let $y$
code the sequence $y_0,\ldots y_{m-1}$, so $[y]_i =y_i$.
Notice that each $x_i$ is distributed
uniformly amongst the four square roots of $[y]_i$.

Run our algorithm, and to each query ``what is the square
root of $[y]_i$?'', answer with $x_i$. 
We will get as output $w$ coding a sequence
$[w]_0, \ldots, [w]_{m-1}$ of square roots of 
$[y]_0, \ldots, [y]_{m-1}$.

If we think of $n$ as fixed, the value of $w$ depends only on the 
inputs given to the algorithm, namely $y$ and the $k$ many
numbers $x_i$ that we gave as replies.
Let $i$ be some index for which $x_i$ was not used. 
Then $x_i$ is distributed
at random among the square roots of $[y]_i$, and $[w]_i$ 
is a square root of $[y]_i$ 
that was chosen without using any information about 
which square root $x_i$ is.
Hence $\gcd(x_i - [w]_i, n)$ 
is a factor of $n$ with probability $1 \over 2$.
\QED
\line

Notice that the only property of the function $|~|$ we used was that we
could find some $n$ with $|n|>k$. So any nondecreasing, not 
eventually constant
function would do in the place of $|~|$. 
Hence if $\PV$ only proves
replacement for very short sequences, that is still enough to 
give us factoring.

In fact under the assumption that factoring is hard we can show that
these replacement schemes form a hierarchy.
For any $\alpha$ with one argument,
let $\BB(\alpha,\PV)$ be the axiom scheme:
\[
\as{i}{\alpha(b)} \es{y}{b} \phi(i,y)
 \implies \E w \as{i}{\alpha(b)} \phi(i,[w]_i)
\]
for all $\PV$ formulas $\phi$. We 
will assume that our
base theory proves that $\alpha(x) <|x|$ and that 
$\alpha$ is increasing.

We need a generalization of a result of Zambella, 
lemma 3.3 of \cite{zambella}. 
The lemma there is presented for a two-sorted system similar to $V^0$
and with $|x|$ rather than $\alpha(x)$.

An $\EPV$ formula is a $\PV$ formula preceded by a bounded
existential quantifier; modulo $\PV$ this is the
same as a strict $\Sb_1$ formula.

\begin{lemma} \label{lem:alphareplacement}
Any model $N \models \PV$ has an $\EPV$-elementary extension to
a model $M \models \PV + \BB(\alpha,\PV)$ such that every element
of $M$ is of the form $f(a,\bb)$ for some $f \in \PV$, $a \in N$ and
$\bb \ss \alpha(M)$, where 
$\alpha(M)=\{ x \in M : x < \alpha(y), \textrm{~some~} y \in M \}$. Informally,
$M$ is formed from $N$ by only adding new ``$\alpha$-small'' elements
and closing under $\PV$ functions.
\QED
\end{lemma}

\Proof
Let $L$ be the language of $\PV$ with the addition of a name for every
element of $N$, and let $T$ be the universal theory of $N$ in this
language, so every model of $T$ will be an $\exists$-elementary,
and hence $\EPV$-elementary, extension of $N$.
Enumerate as $( t_1, \phi_1(x,y) ),
( t_2, \phi_2(x,y) ), \ldots$ all pairs consisting
of closed terms in $L$ and binary $\PV$ formulas with parameters from $L$.
We will use this to construct a chain $T=T_0 \ss T_1 \ss T_2 \ss \ldots$
of theories.

Suppose that $T_i$ has been constructed and is a consistent, universal 
theory. 
If $T_i \proves \as{x}{\alpha(t_{i+1})} \E y \phi_{i+1}(x,y)$
then put $T_{i+1}=T_i$. Otherwise introduce a new constant symbol $c$ and
put
\[
T_{i+1} = T_i \cup \{ c<\alpha(t_{i+1}) \} \cup \{ \A y \neg \phi_{i+1} (c,y) \}.
\]
Note that $T_{i+1}$ is consistent and universal.

Let $T^*$ be the union of this chain of theories, and let $L^*$ be $L$ 
together with all the new constant symbols that were added in the construction
of $T^*$. Enumerate all pairs of closed terms and binary formulas in $L^*$, and
repeat the above construction to get a theory $T^{**}$ and a language $L^{**}$.
Repeat this step $\omega$ times, and let $T^+$ be the union of the theories
and $L^+$ its language.

$T^+$ is consistent and universal, so there is a model $M \vDash T^+$
each element of which is named by some closed $L^+$-term.
$M \vDash T$, so $M$ is an $\EPV$-elementary extension of $N$.
Also, each time a new constant $c$ was introduced to $L^+$, 
$c<\alpha(t)$ was introduced to $T^+$ for some term $t$. So
$M$ is the closure of elements of $N$ and new ``$\alpha$-small''
elements, as required.

To show that $M$ is a model of $\BB(\alpha,\PV)$, suppose that $a$ is an
element of $M$ and $\phi(x,y)$ is a $\PV$ formula with parameters
from $M$, and 
\[
M \vDash \as{x}{\alpha(a)} \E y \phi(x,y).
\]

Then by the construction of $M$, we may assume that $a$ is named
by some closed $L^+$ term $t$ and that $\phi(x,y)$ is a parameter-free 
$L^+$ formula; and by the construction of $T^+$ we must have that
$T^+ \proves \as{x}{\alpha(t)} \E y \phi(x,y)$,
since $T^+$ either proves this or its negation.
But $T^+$ is a universal theory, so by using Herbrand's theorem
and the properties of $\PV$ we can find a $\PV$
function symbol $f$ (with parameters) such that
$T^+ \proves \as{x}{\alpha(t)} \phi(x,f(x))$.
Now by the comprehension available in $\PV$, we can find
some $w \in M$ such that $M \vDash \as{x}{\alpha(t)} \phi(x,[w]_x)$,
as required.
\QED
\line

We can now adapt the proof of the KPT witnessing theorem
to get the following:

\begin{theo} \label{the:BBalpha}
Suppose
$$
\PV + \BB(\alpha,\PV) \proves \A x \E y \A z \phi(x,y,z)
$$
for an $\EPV$ formula $\phi$. Then there exist $k \in \N$, a term 
$s(x,\bar{z})$ and
functions $f_1, \ldots, f_k$ such that 
\begin{align*}
\PV \proves  &  \A x \A {\bar{z},}
 \es{i}{\alpha(s)^k}\phi(x,[f_1(x)]_i,[z_1]_i) \\
& \disj
\es{i}{\alpha(s)^k}\phi(x,[f_2(x,z_1)]_i,[z_2]_i) \\
& \disj \ldots \disj 
\es{i}{\alpha(s)^k}\phi(x,[f_k(x,z_1,\ldots,z_{k-1})]_i,[z_k]_i)
\end{align*}
(we include the exponent $k$ here because the range of $\alpha$ 
might not be closed under multiplication).
\end{theo}

\Proof
Enumerate all pairs of $\PV$ functions as $(s_1,f_1), (s_2,f_2), \ldots$
with infinite repetitions in such a way that for each $k$ both
$s_k$ and $f_k$ take $k$ or fewer arguments.
Assume that the conclusion of the theorem is false, and let
$T$ be the theory
\begin{align*}
\PV + &   \{
\as{i}{\alpha(s_1(b,c_1))^1}  \neg \phi(b,[f_1(b)]_i,[c_1]_i),\\
& \as{i}{\alpha(s_2(b,c_1,c_2))^2} \neg \phi(b,[f_2(b,c_1)]_i,[c_2]_i),
\ldots \}
\end{align*}
where $b$ and $c_1,c_2, \ldots$ are new constant symbols. Then 
$T$ is finitely satisfiable (we can take the term $s$ in the statement
of the theorem as the sum of our finite set of terms $s_1, \ldots, s_k$).

Let $N$ be a model of $T$, and let $N' \ss N$ be the substructure 
consisting of all the elements named by terms. Since $T$ is universal, 
$N' \models T$. Let $M$ be the extension of $N$ given by lemma
\ref{lem:alphareplacement} to a model of $\BB(\alpha,\PV)$. By
$\EPV$ elementariness, $M$ is also a model of $T$. 

Now let $a$ be any element of $M$. By the construction of $M$, 
for some $\bar{d} \ss \alpha(M)$, some $e \in N'$ and some $\PV$ function $g$
we have $a=g(\bar{d},e)$. Furthermore by the construction of $N'$ we know
that $\bar{d}<\alpha(h_1(b,c_1, \ldots, c_k))$ and 
$e=h_2(b,c_1, \ldots, c_k)$ for some $k$ and some $\PV$ functions $h_1$
and $h_2$.

In this paragraph we identify a number $i<\alpha(h_1(b,\bc))^k$ 
with the sequence $\bar{i}=i_1 \ldots i_k$
of numbers less than $\alpha(h_1(b,\bc))$ that it codes. 
We can find $l > k$ such that $f_l$ is the $\PV$ function symbol
that takes as input $b,c_1,\ldots,c_l$ and outputs (as a single number)
the sequence $w_1 \ldots w_{\alpha(h_1(b,c_1, \ldots, c_k))^k}$ where
$w_i = g(\bar{i},h_2(b, c_1, \ldots, c_k))$.  
Then $a=[f_l(b,c_1,\ldots,c_l)]_d$ and since $M \models T$
we have $M \models \neg \phi(b,a,[c_{l+1}]_d)$. Here
$a$ was chosen arbitrarily, so we have shown that 
$M \models \PV + \BB(\alpha,\PV) + \neg \A x \E y \A z \phi(x,y,z)$.
\QED

\begin{cor}
Suppose that factoring is not possible in probabilistic polynomial
time. Then
$\BB(\alpha, \PV)$ is not provable in $\PV + \BB(\beta,\PV)$,
for terms $\alpha, \beta$ where 
$\alpha(x),\beta(x)<|x|$ and
$\alpha$ grows 
faster than any polynomial in $\beta$.
\end{cor}

\Proof
Our standard argument is that if replacement is provable
in $\PV$, then there is a polynomial time
interactive algorithm that queries
$k$ square roots and outputs $|n|$ square roots, for some
fixed $k \in \N$.

By theorem \ref{the:BBalpha} we can show, by a similar
argument, 
that if $\PV + \BB(\beta,\PV) \proves \BB(\alpha, \PV)$
then we have a polynomial time interactive algorithm that queries
$k \beta(n)^k$ square roots modulo $n$ and outputs $\alpha(n)$ 
square roots, for some fixed $k \in \N$.

So if $n$ is sufficiently large that 
$\alpha(n) > k \beta(n)^k$, we can use the argument
of theorem \ref{the:roots} to factor $n$.
\QED
\line

This gives a hierarchy of theories 
\[
\PV + \BB(|x|,\PV) \supset \PV + \BB(||x||,\PV) \supset \ldots
\]

The same argument goes through in $V^0$. 
One way to see this is to notice
that the important difference between $\PV$ and $V^0$ is that
the $\PV$ functions are closed under polynomial time iteration,
and no such iteration is used in the proof here. So we have
the unconditional separation result

\begin{theo}
$\BB(\alpha, \SB)$ is not provable in 
$V^0 + \BB(\beta,\SB)$, for
 terms $\alpha, \beta$ where 
 $\alpha(n),\beta(n)<n$ and
 $\alpha$ grows 
faster than any polynomial in $\beta$.
\end{theo}

\Proof
If the theorem
is false, then there is $k \in \N$ and an interactive
algorithm that, given $\alpha(n)$
many vectors $v_1, \ldots, v_{\alpha(n)}$,
each of length $n$, will make $k\beta(n)^k$
queries of the form ``what is the parity vector of $v_i$?''
and then output the parity vectors of all the $v_i$s.
So if $\alpha(n) \ge 3 k \beta(n)^k$, then by adapting
the argument of section 3 we get a probabilistic 
uniform $\AC0$
algorithm which computes parity.
\QED

\section{Unique replacement in PV and RSA}

We define ``unique replacement'' to be the scheme
\[
\as{i}{|a|}\es{!x}{b}\phi(i,x) 
 \implies \E w \as{i}{|a|}\phi(i,[w]_i).
\]

\begin{theo} \label{the:RSA}
If $\PV$ proves unique replacement for sharply bounded formulas,
then the injective $\WPHP$ for $\PV$ formulas can be witnessed in 
probabilistic polynomial time 
(and hence in particular we can crack RSA \cite{kpcrypto}).
\end{theo}

\Proof (Simplified from the model-theoretic
proof in \cite{thapen}.)
First notice that it is sufficient to show that $\PV$ does not 
prove unique replacement for some $\PV$ formula $\phi$. For
suppose that $\phi$ is decided by the polynomial time machine with code $e$,
and that for some fixed $i$
there is a unique $x$ such that $\phi(i,x)$. Then there 
is a unique pair $(z,x)$ such that $z$ is an accepting computation 
of the machine $e$ on input $(i,x)$, and the property of being an
accepting computation is sharply bounded.

In the rest of this proof $x$ and $y$ will code sequences
of $|n|$ numbers each of size $<n^{|n|}$ 
and with elements $[x]_i, [y]_i$, and 
$z$ will code  
a sequence of $|n|$ numbers each of size $<n$ and with elements $\ang{z}_i$.

Suppose that $h$ is a $\PV$ function from $n^{|n|}$ to $n$.
Note that from any $\PV$ function $g:2n \rightarrow n$ we can derive 
such a function $h$ with the property that a witness to $\WPHP$ for $h$ 
yields in polynomial time a witness to $\WPHP$ for $g$ (\cite{wilkiepidge},
or see \cite{thapen} for an explicit polynomial time construction).

Choose $x < n^{|n|^2}$ at random and let 
$z<n^{|n|}$ be such that
$\ang z_0 = h([x]_0), \ldots, \ang z_{|n|-1}=h([x]_{|n|-1})$.

Assume that $\PV$ proves the following instance of unique replacement:
\begin{align*}
& \es{i}{|n|}\as{u}{n^{|n|}}h(u) \neq \ang z_i \\
& \disj
\es{i}{|n|}\es{u_1 \! < \! u_2}{n^{|n|}} h(u_1)=h(u_2) \\
& \disj
\es{y}{n^{|n|^2}}\as{i}{|n|} h([y]_i)=\ang z_i.
\end{align*}

Then by our witnessing theorem, for some $k$ (independent
of $n$) there is a deterministic interactive computation which takes
$n$ and $z$ as its initial input. Then for $k$ steps it gives us an
index $i<{|n|}$ and expects an input $y<n^{|n|}$;
if we can guarantee that
for each such step we have $h(y)=\ang z_i$, then the computation outputs
either $u_1$ and $u_2$ mapping to the same thing, in which case we
are done (and this case is the only one that is different from 
normal replacement), or 
$y<n^{|n|^2}$ satisfying $\as{i}{|n|} h([y]_i) =\ang z_i$.

Run the computation, and to each index $i$ queried respond with $[x]_i$.
The computation must output some $y$ satisfying 
$\as{i}{|n|} h([y]_i) = \ang z_i$. 
 Now the computation is deterministic,
and if we think of $n$ as fixed, there were $n^{|n|(k+1)}$ possible different
inputs to the machine: namely $n^{|n|}$ different possibilities for $z$
and $(n^{|n|})^k$ different possibilities for the $k$ responses $[x]_i$.
Hence there are at most $n^{|n|(k+1)}$ possible outputs $y$.
However $x$ was originally chosen at random from $n^{|n|^2}$ possibilities. 
So if $k<n-1$ then with high probability $x$ is not a possible
output of the machine, so 
$x \neq y$ and for some $i<|n|$
we have $[x]_i \neq [y]_i$ but $h([x]_i)=\ang z_i=h([y]_i)$.
\QED
\line

Notice that part of this argument can be formalized in $\PV$, to show
that if $\PV$ proves unique replacement, then $\PV$ proves
that the surjective $\WPHP$ for $\PV$ functions implies the injective
$\WPHP$ for $\PV$ functions. In the proof above randomness was used
to find some $x$ outside the range of a given polynomial time algorithm;
in the formal $\PV$ proof we would use the surjective $\WPHP$ to 
provide such an $x$.

\begin{cor}
Suppose
 $\PV$ proves the $\Db_1$ comprehension axiom scheme (\ref{compre}).
Then 
$\PV$ proves unique replacement for $\PV$ formulas and
by theorem \ref{the:RSA} we can crack RSA.
\end{cor}

\Proof
Let $\phi(i,x)$ be any $\PV$ formula (with parameters) and
suppose that the hypothesis of the theorem holds.
Let $M \models
\PV$, $a,b \in M$ and suppose $M \models \as{i}{|b|} \es{!x}{a} \phi(i,x)$.
Then 
\begin{align*}
M &  \models \as{i}{|b|} \as{j}{|a|,}  \\
 & \es{x}{a}(\phi(i,x) \conj x_j = 1) \leftrightarrow
 \as{x}{a}(\phi(i,x) \implies x_j = 1).
\end{align*}

Over $\PV$, $\phi$ is equivalent to both a $\Sb_1$ and a $\Pb_1$
formula, so we can apply 
comprehension and get some $w$ such that
\begin{align*}
M \models  \as{i}{|b|} & \as{j}{|a|,} \\
 &   ([w]_i)_j = 1 
\leftrightarrow  \es{x}{a}(\phi(i,x) \conj x_j = 1).
\end{align*}
Here we assume without loss of generality that $a$ is a power of $2$,
so that we can switch easily between thinking of $w$ as a binary 
sequence of length $|b||a|$ and as a sequence of $|b|$ many binary numbers
$[w]_1 \ldots [w]_{|b|},$
each of length $|a|$. 
We also use the fact that in $\PV$ the
formula $\phi(i,x)$ can be written in both a strict $\Sb_1$ and a
strict $\Pb_1$ way, which we need to apply comprehension.

Now pick any $i<|b|$. There is some unique $x \in M$ such that 
$\phi(i,x)$;
and by the construction of $w$, for each $j < |a|$ we know $([w]_i)_j=1$ 
if and only if $x_j=1$.
Hence $[w]_i=x$.

So $M \models \as{i}{|b|} \phi(i,[w]_i)$.
\QED

\bibliographystyle{plain}
\bibliography{lics}

\end{document}